\documentclass[onecolumn,preprint]{elsarticle}

\usepackage{citesort}
\usepackage{lipsum}
\usepackage{mathtools}
\usepackage{cuted}
\usepackage{graphicx}

\begin{document}
\title{The ORNL Analysis Technique for Extracting $\beta$-Delayed Multi-Neutron Branching Ratios with BRIKEN}

\author{B.C. Rasco}
\ead{brasco@utk.edu}
\author{N.T. Brewer}
\address{Physics Division, Oak Ridge National Laboratory, Oak Ridge, TN, 37831-6371, USA}
\address{JINPA, University of Tennessee at Knoxville, Knoxville, TN, 37831, USA}

\author{R. Yokoyama}
\author{R. Grzywacz}
\address{Department of Physics and Astronomy, University of Tennessee at Knoxville, Knoxville, TN, 37966, USA}

\author{K.P. Rykaczewski}
\address{Physics Division, Oak Ridge National Laboratory, Oak Ridge, TN, 37831-6371, USA}

\author{A. Tolosa-Delgado}
\author{J. Agramunt}
\author{J.L. Ta{\'i}n}
\address{Instituto de Fisica Corpuscular (CSIC-Universitat de Valencia), E-46071 Valencia, Spain}

\author{A. Algora} 
\address{Instituto de Fisica Corpuscular (CSIC-Universitat de Valencia), E-46071 Valencia, Spain}
\address{Institute for Nuclear Research (MTA Atomki), H-4001 Debrecen, POB.51., Hungary}

\author{O. Hall}
\author{C. Griffin}
\author{T. Davinson}
\address{University of Edinburgh, EH9 3JZ Edinburgh, U.K.}

\author{V.H. Phong}
\address{VNU University of Science, Hanoi, Vietnam}
\address{RIKEN Nishina Center, 2-1 Hirosawa, Wako-shi, Saitama 351-0198, Japan}
\author{J. Liu}
\address{The University of Hong Kong, Hong Kong, China}
\address{RIKEN Nishina Center, 2-1 Hirosawa, Wako-shi, Saitama 351-0198, Japan}
\author{S. Nishimura}
\address{RIKEN Nishina Center, 2-1 Hirosawa, Wako-shi, Saitama 351-0198, Japan}

\author{G.G. Kiss}
\address{RIKEN Nishina Center, 2-1 Hirosawa, Wako-shi, Saitama 351-0198, Japan}
\address{MTA Atomki, Bem t{\'e}r 18/c, Debrecen H4032, Hungary}

\author{N. Nepal}
\address{Science of Advanced Materials Program, Central Michigan University, Mount Pleasant, MI, 48859, USA}

\author{A. Estrade}
\address{Department of Physics, Central Michigan University, Mount Pleasant, MI, USA}



\begin{abstract}

Many choices are available in order to evaluate large radioactive decay networks. 
There are many parameters that influence the calculated $\beta$-decay delayed single and multi-neutron emission branching fractions.
We describe assumptions about the decay model, background, and other parameters and their influence on $\beta$-decay delayed multi-neutron emission analysis.
An analysis technique, the ORNL BRIKEN analysis procedure, for determining $\beta$-delayed multi-neutron branching ratios in $\beta$-neutron precursors produced by means of heavy-ion fragmentation is presented.
The technique is based on estimating the initial activities of zero, one, and two neutrons occurring in coincidence with an ion-implant and $\beta$ trigger.
The technique allows one to extract $\beta$-delayed multi-neutron decay branching ratios measured with the hybrid \textsuperscript{3}He BRIKEN neutron counter.
As an example, two analyses of the $\beta$-neutron emitter \textsuperscript{77}Cu based on different {\it a priori} assumptions are presented along with comparisons to literature values.
\end{abstract}


\maketitle 

\section{Introduction}
Measuring single and multi-neutron emission after $\beta$ decay of neutron-rich nuclei is important in order to understand the evolution of nuclear structure and its impact on $\beta$-decay properties far from stability.
Multi-neutron emission after $\beta$ decay of neutron-rich nuclei also impacts astrophysical r-process calculations that estimate the abundance of various nuclei in the galaxy \cite{Mumpower2012a, Surman2015}.
Present and future $\beta$-decay experiments with neutron-rich exotic nuclei created from the fragmentation of heavey ions involve complex decay networks.
It is important to have a robust method to reliably extract the decay information associated with each nucleus.
The $\beta$ delayed neutrons at RIKEN (BRIKEN) collaboration measured the $\beta$ decays of many neutron-rich nuclei that exhibit zero, single, and multi-neutron emission probabilities, $P_{xn}$ (where $x=0,1,2,...$) \cite{Tain2018}. 

Techniques for evaluating single neutron branching ratios, $P_{1n}$, with \textsuperscript{3}He tubes \cite{belen2016,Gomez2011} must be extended to include the possibility of multi-neutron $\beta$ decay.
So far, in heavy nuclei, only one case of a large $\beta$-delayed 2 neutron emitter, \textsuperscript{86}Ga ($P_{2n}=20(10)$\%), has been reported \cite{Miernik2013}.
The BRIKEN collaboration aims to extend current knowledge of two and more neutron emitters in medium and heavy mass nuclei \cite{Tain2018}.

In this paper, we present a technique based upon estimating the initial activities of zero, one, and two neutrons detected in coincidence with an ion-implant and a $\beta$ trigger.
There are several challenges evaluating $P_{xn}$ and the associated systematic and statistical uncertainties.
This paper discusses these challenges and presents one analysis procedure, the ORNL BRIKEN analysis procedure, used to evaluate $P_{xn}$. 
Alternative analysis methods will be published separately.
As an example, two analyses of \textsuperscript{77}Cu BRIKEN data are presented and compared with previous measurements.

\section{BRIKEN Detector Description}

The hybrid BRIKEN detector consists of 140 \textsuperscript{3}He neutron detector tubes, a dual purpose ion-implant and $\beta$ detector (implant-$\beta$ detector), and two HPGe clovers.
The BRIKEN detector was designed to maximize the neutron efficiency while keeping the neutron efficiency as uniform as possible over a wide range of initial neutron energies. 
The uniform neutron efficiency minimizes the contribution to the neutron efficiency uncertainty from the initial neutron kinetic energy.
This effect and its impact on the BRIKEN design is discussed in \cite{tar2017}.
From the analysis presented in \cite{tar2017} and neutron source measurements, the average single neutron efficiency of the BRIKEN detector is 62(2)\% for neutrons with kinetic energies ranging from thermal energies to 5 MeV.

BRIKEN was placed on the zero degree beam line following BigRIPS at the RI Beam Factory (RIBF) of the RIKEN Nishina Center.
The nuclei were identified by means of the BigRIPS separator \cite{Kubo2003}.

Several different implant-$\beta$ detectors were used in the various BRIKEN experimental runs at RIKEN. 
Two different silicon based implant-$\beta$ detectors were used in separate runs, the AIDA detector \cite{Griffin2014} and the WAS3ABi detector \cite{Nishimura2013}. 
In conjunction with the WAS3ABi detector, a YSO scintillator \cite{Balcerzyk2000} based implant-$\beta$ detector was also used.
All of the implant-$\beta$ detectors are segmented in order to reduce ion-correlated background $\beta$ triggers.
Two HPGe clovers from the CLARION array of Oak Ridge National Laboratory were used to detect $\gamma$ rays in coincidence with $\beta$ and $\beta$-delayed neutron decays.

The present paper discusses the analysis of BRIKEN data using as an example \textsuperscript{77}Cu data that was taken with the AIDA implant detector. 
The analysis of \textsuperscript{77}Cu is chosen because it is a known $\beta$-delayed neutron emitter, with a half life of $468(2)$ ms \cite{Wang2016} and a consistently measured single neutron decay fraction, $P_{1n}=31.0(38)\%$ \cite{hosmer2009} and $P_{1n}=30.3(22)\%$ \cite{Ilyushkin2009}.
The present paper does not comment on the evaluation of the associated $\gamma$-ray detection, which will be presented in a future publication.
Before presenting the ORNL BRIKEN analysis method, we offer comments on the inputs and parameters and the sources of errors in evaluating $P_{xn}$.

\section{Connecting Activities Gated on Neutron-Multiplicity to $P_{xn}$} 

Calculating $P_{xn}$ involves evaluating the number of correlated implant triggers with $\beta$ triggers versus implant-$\beta$ times ($\beta$ time minus implant time), hereafter referred to as implant-$\beta$ activities.
Using the estimated initial activity (the activity at the implant time) from the implant-$\beta$ activity gated in coincidence on the neutron multiplicity gives a way to obtain the $P_{xn}$.

\begin{figure}[p!]
\begin{center}
\includegraphics[width=1.0\linewidth]{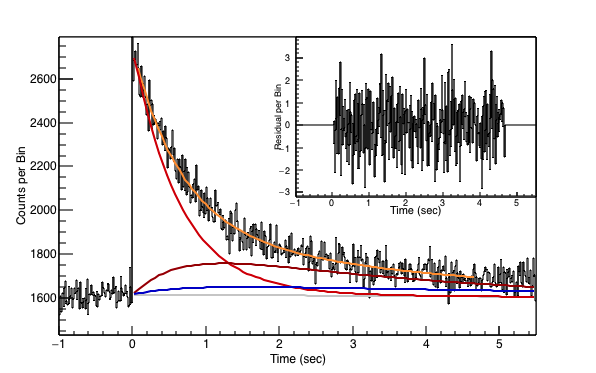}
\caption
{
Fit of adapted Bateman equation to \textsuperscript{77}Cu data with an implant-$\beta$ trigger correlation and no information on the number of neutrons from the \textsuperscript{3}He tubes.
The total fit is shown in orange, \textsuperscript{77}Cu is shown in red, \textsuperscript{77}Zn is shown in dark red, \textsuperscript{76}Zn is shown in blue, the background is shown in gray, and the data are shown in black.
All decay curves are offset by the background.
The granddaughter decays are not shown to preserve clarity.
}
\label{fig:cu77_total_mod_bateman_fit}
\end{center}
\end{figure}

\begin{figure}[p!]
\begin{center}
\includegraphics[width=1.0\linewidth]{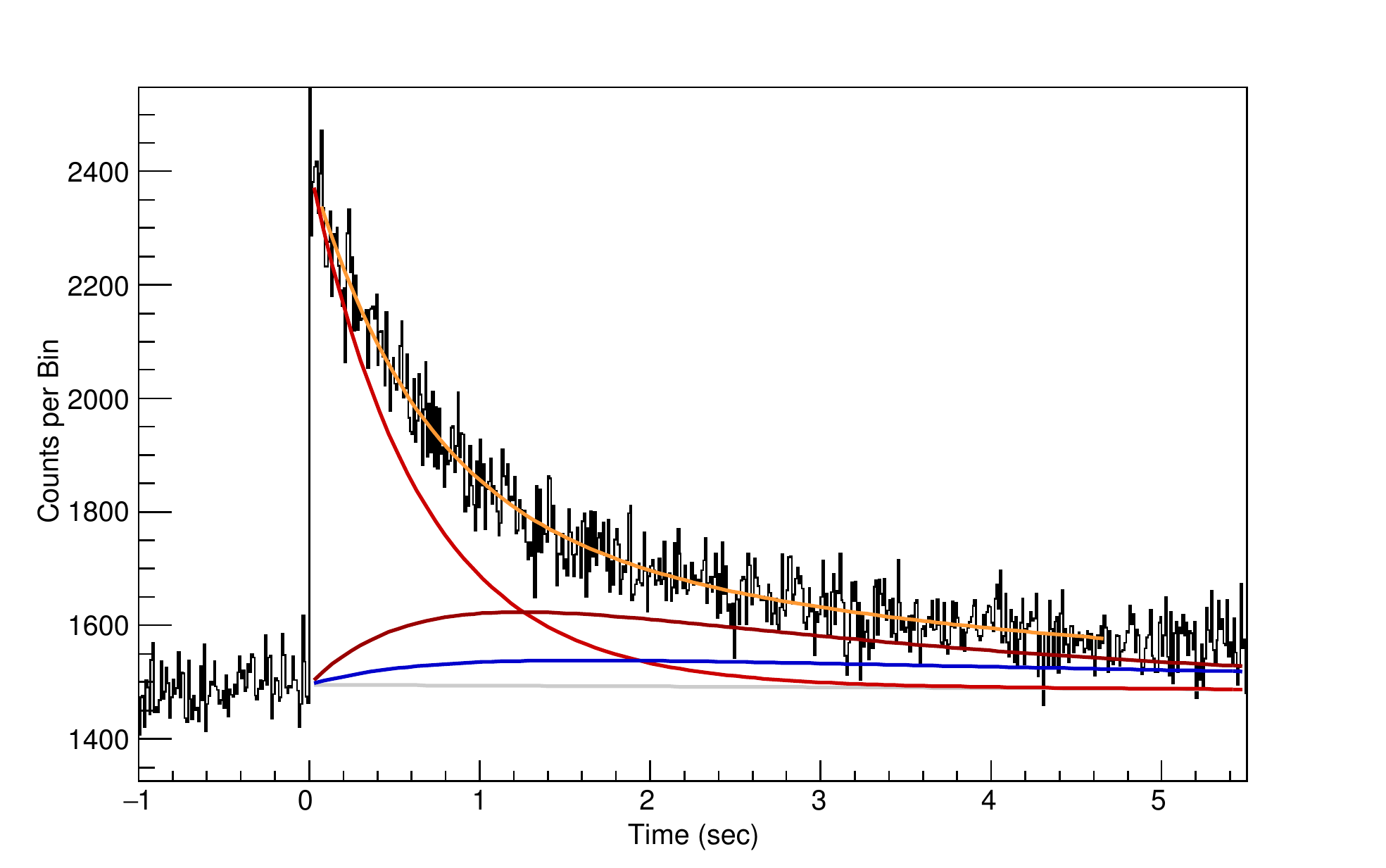}
\caption{
Fit of adapted Bateman equation to \textsuperscript{77}Cu data with an implant-$\beta$ trigger correlation and zero neutrons detected in the \textsuperscript{3}He tubes.
Colors and comments are as in figure \ref{fig:cu77_total_mod_bateman_fit}.
}
\label{fig:cu77_0n_fit}
\end{center}
\end{figure}

\begin{figure}[p!]
\begin{center}
\includegraphics[width=1.0\linewidth]{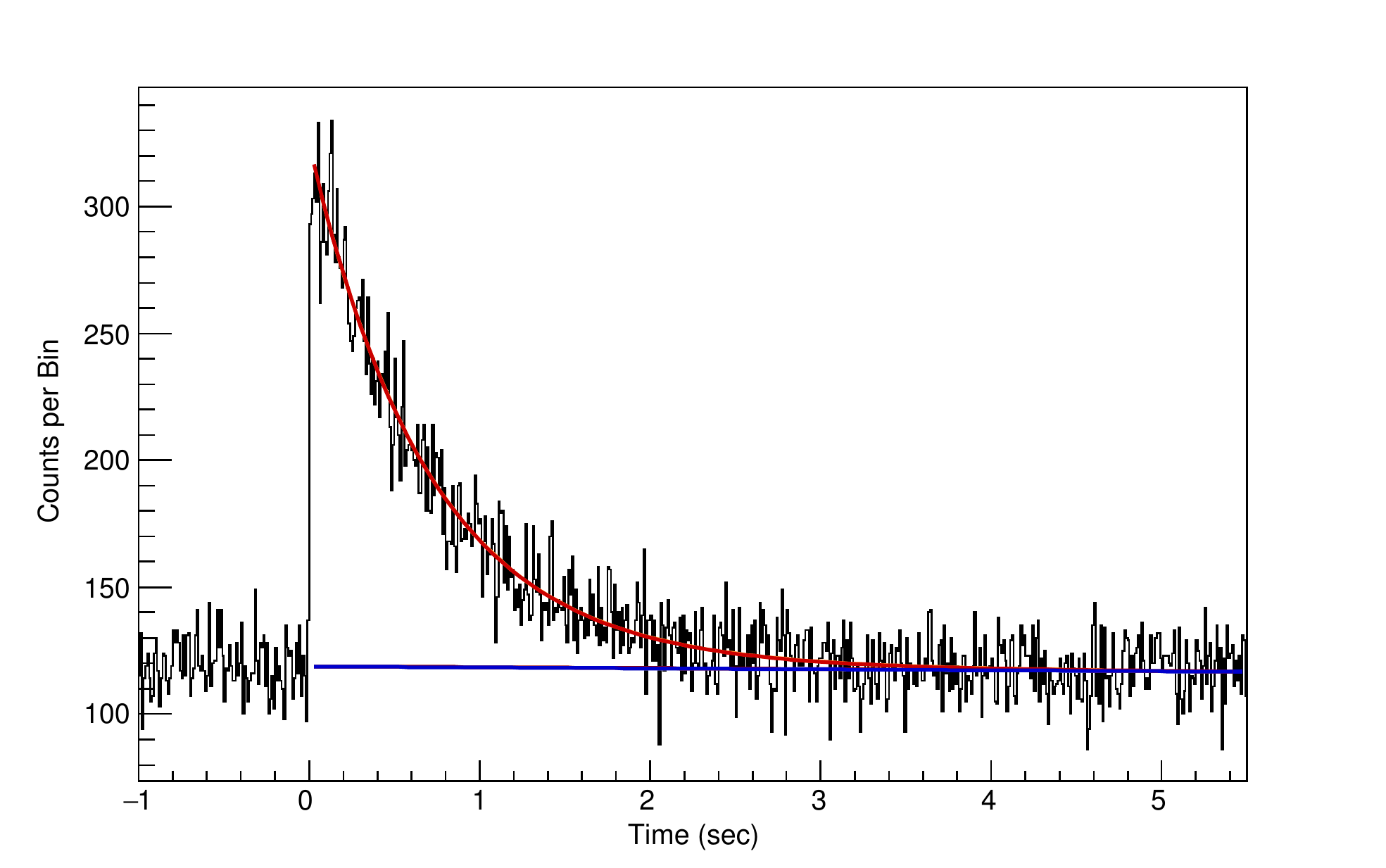}
\caption{
Fit of the adapted Bateman equation to \textsuperscript{77}Cu data with an implant-$\beta$ trigger correlation and one neutron detected in the \textsuperscript{3}He tubes.
Colors and comments are as in figure \ref{fig:cu77_total_mod_bateman_fit}, though the total and the \textsuperscript{77}Cu decay are indistinguishable.
}
\label{fig:cu77_1n_fit}
\end{center}
\end{figure}

\begin{figure}[p!]
\begin{center}
\includegraphics[width=1.0\linewidth]{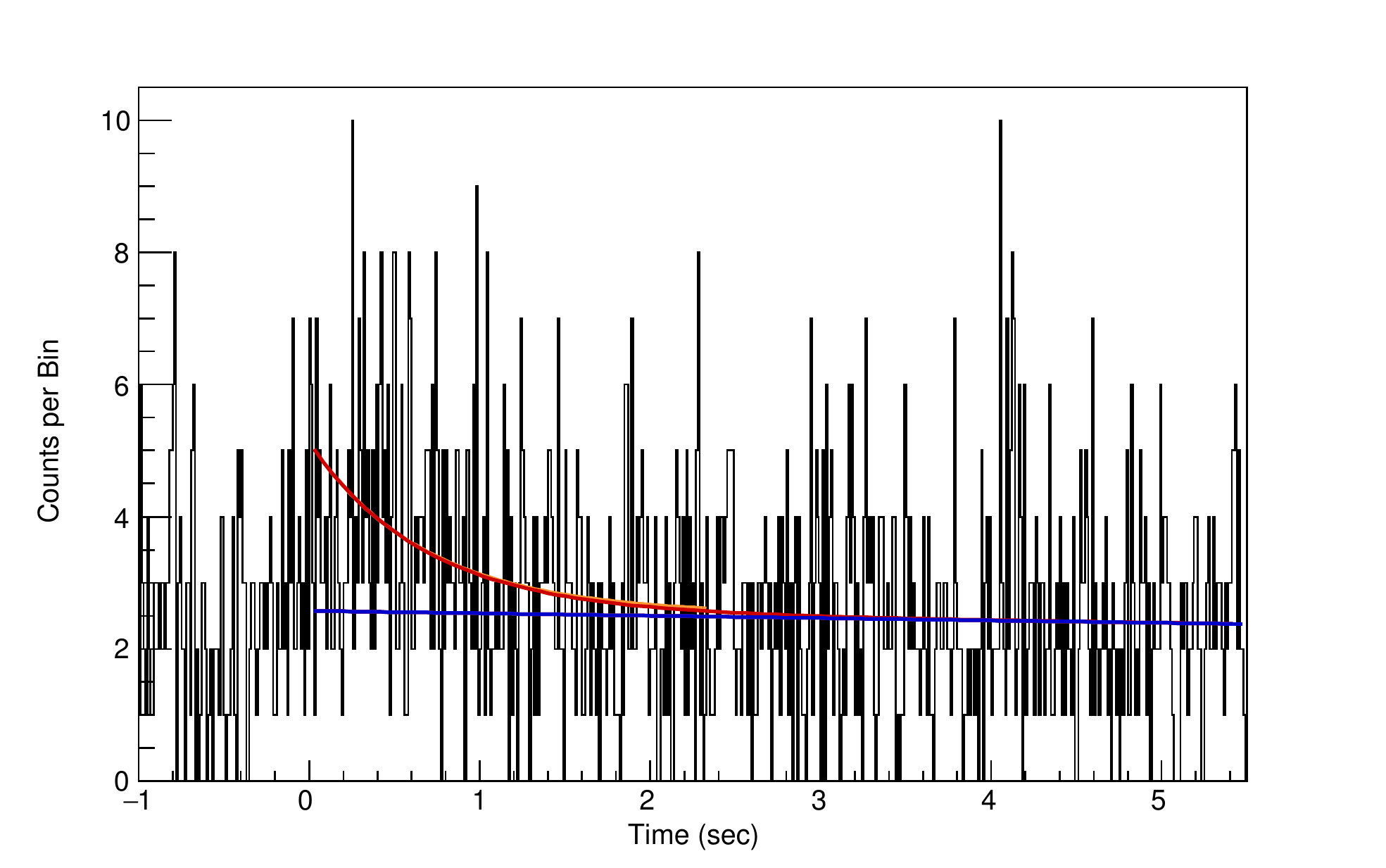}
\caption{
Fit of the adapted Bateman equation to \textsuperscript{77}Cu data with an implant-$\beta$ trigger correlation and two neutrons detected in the \textsuperscript{3}He tubes.
Colors and comments are as in figure \ref{fig:cu77_total_mod_bateman_fit}, though the total and the \textsuperscript{77}Cu decay are indistinguishable.
}
\label{fig:cu77_2n_fit}
\end{center}
\end{figure}

For each ion-implant signal all associated $\beta$ signals within $\pm 10$ sec within $\pm 3$ pixels of the implant pixel of AIDA are correlated in software. 
Each pixel in AIDA has a 0.58 mm pitch in both the x and y direction.
The implant-$\beta$ time correlation plot from a 60 hour BRIKEN run for BigRIPS selected \textsuperscript{77}Cu implanted ions is shown in figure \ref{fig:cu77_total_mod_bateman_fit}.
In addition to the implant-$\beta$ time correlation activity plots, there are implant-$\beta$ time correlation activity plots gated on the number of neutrons detected within the neutron thermalization time window, $T_{th}=200$ $\mu$s, after each $\beta$ signal (neutron-multiplicity implant-$\beta$ activities).
The activity gated on zero neutrons detected is shown in figure \ref{fig:cu77_0n_fit}, the activity gated on one neutron detected is shown in figure \ref{fig:cu77_1n_fit}, and the activity gated on two neutrons detected is shown in figure \ref{fig:cu77_2n_fit}.
Below we describe how the estimated initial activity of the neutron-multiplicity implant-$\beta$ activities are used to calculate the $P_{xn}$.

Before discussing the connections between the initial activity of the neutron-multiplicity implant-$\beta$ activities and the $P_{xn}$, a discussion of several required parameters is presented.
Some of these required parameters can be measured, while others must be estimated.
The evaluation and propagation of uncertainties from measured and estimated parameters through the analysis is presented.
A discussion of the parameters considered in the BRIKEN $P_{xn}$ evaluations is given below.

\subsection{Implant-$\beta$ Background}

Random $\beta$ signals in coincidence with each implant contribute to the nearly constant background in each implant-$\beta$ time correlation plot.
These random $\beta$ signals originate from other nearby implant $\beta$ signals and implant $\beta$ signals that are not detected by the $\beta$ trigger. 
The small slope of the background is associated with short time drops (up to tens of seconds) in the rate of implanted ions from an otherwise DC beam.
When the beam drops before an implant, this lowers the correlated $\beta$ counts before the implant.
Similarly, beam drops after an implant lower the background counts after the implant.
Because there are relatively few beam drops, this is a small yet observable effect.

An accurate description of the background affects the fitting of the neutron-multiplicity implant-$\beta$ activities.
Especially when the background models differ on the order of the daughter and granddaughter activities.
One way to minimize the impact of the background modeling is to fit over a shorter time, this minimizes the impact of variations of the background.
For the \textsuperscript{77}Cu zero neutron-multiplicity implant-$\beta$ activity, the background slope is on the order of $1.5$ counts per second, while for the \textsuperscript{77}Cu one neutron-multiplicity implant-$\beta$ activity, the background slope is on the order of $0.2$ counts per second.
While this is small, it contributes a bias to the fit of the \textsuperscript{77}Cu descendent activities.

The background is linearly modeled, $C_{0}+C_{1}*t$, before the implant and it is assumed that the background after the ion-implant time is linearly modeled as, $C_{0}-C_{1}*t$, with $C_{0}$ and $C_{1}$ calculated from the background before the implant.
There is some uncertainty in this assumption and an approach is taken to minimize the impact of the background uncertainty on the estimation of the initial activity.

The ion-implants have very little background signal, due to the large unique signal of stopping a heavy ion with $100-200$ MeV/u energy and the isotopic identification plus coincident timing from the BigRIPS detectors \cite{Kubo2003}, though the ion-implants do create background in the other detectors.

\subsection{\textsuperscript{3}He Neutron Detector}

The neutron-rich nuclei studied have roughly $100-200$ MeV/u of kinetic energy and their implantation creates background signals in all of the detectors, including the silicon, scintillator, $\gamma$, and \textsuperscript{3}He neutron detectors.
The \textsuperscript{3}He detectors see two types of background neutron counts. 
The first type of background the \textsuperscript{3}He counters see is an increase in neutron and $\gamma$ counts associated with the implanted energetic ion, referred to as the prompt flash.
The second type of neutron counter background is from the neutron room background in online conditions, referred to as random neutron background.

The prompt flash neutron background associated with the stopping of energetic ions detected in the \textsuperscript{3}He counters is removed by rejecting neutrons detected in the \textsuperscript{3}He counters within one neutron thermalization time, $T_{th}$, after the implant time.

Random neutron backgrounds contribute to the implant-$\beta$ activities time structure since they occur in coincidence with the $\beta$ signal, and therefore these need to be accounted for in the analysis.
Random neutron background probability coincidences that occur within one neutron thermalization time window after the $\beta$-trigger time in the \textsuperscript{3}He detectors are denoted by $r_{0n}$ for the probability of zero background neutrons detected in coincidence, $r_{1n}$ for the probability of one random background neutron detected, and $r_{2n}$ for the probability of two random background neutrons detected within $T_{th}$ of the $\beta$-signal time (written generally as $r_{xn}$ where $x=0,1,2,...$).

\begin{figure}[p!]
\begin{center}
\includegraphics[width=1.0\linewidth]{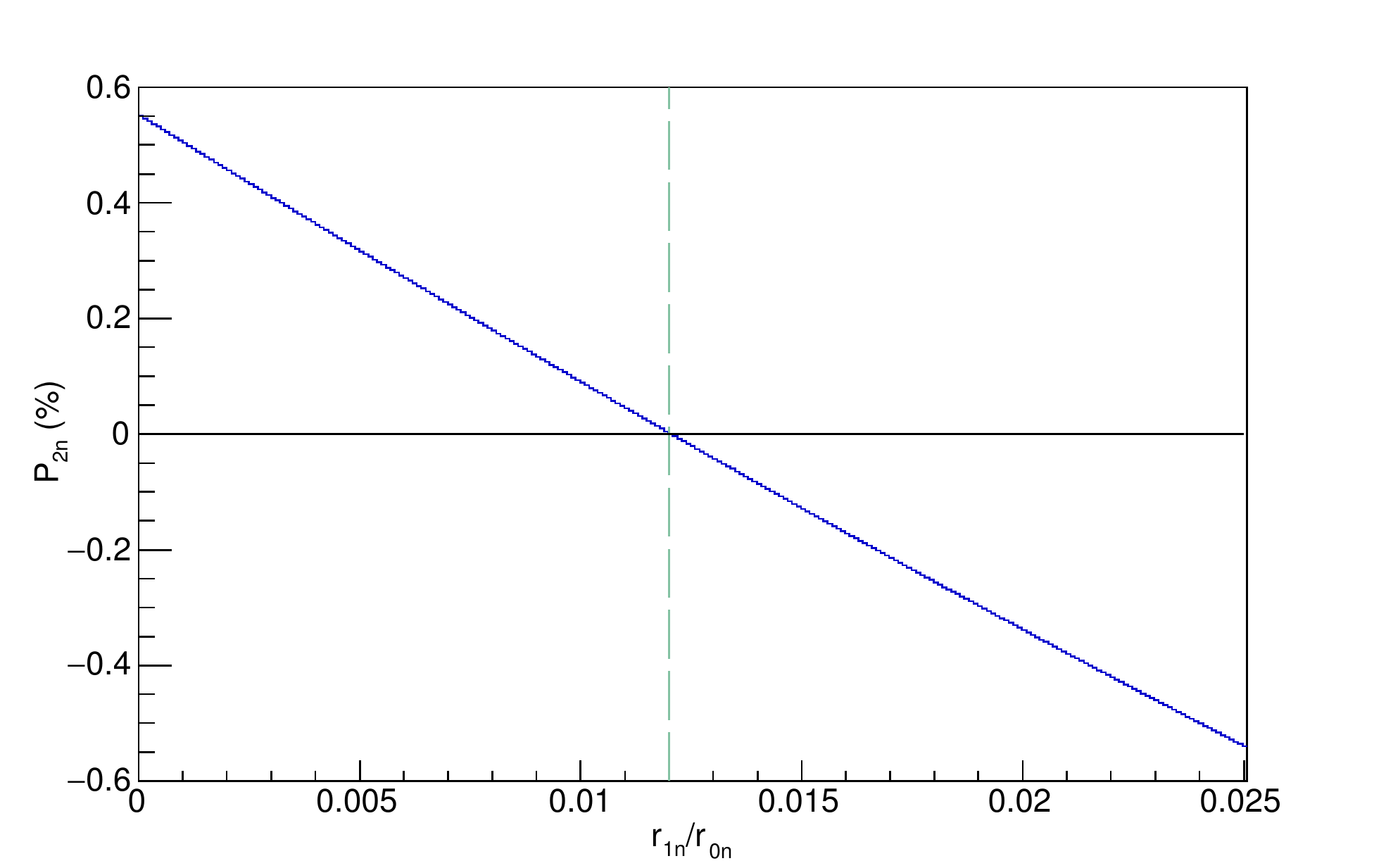}
\caption{
The variation of the calculated \textsuperscript{77}Cu $P_{2n}$ with statistical uncertainties versus the ratio of one neutron background coincidence probability to zero neutron background coincidence probability.
The vertical dashed line at 0.012 is the zero crossing point. 
}
\label{fig:cu77_p2nvsnoise_fit}
\end{center}
\end{figure}

The magnitude of the background neutron coincidence probability, $r_{xn}$, can be estimated by requiring decays that have no possible $P_{2n}$ decay ($Q_{\beta 2n}<0.0$) to have an average calculated $P_{2n}$ consistent with zero. 
This requirement leads to an estimation of the background neutron coincidence probabilities.
Using the analysis presented below, the predicted \textsuperscript{77}Cu $P_{2n}$ versus the ratio of the probability of detecting one neutron to detecting zero neutrons, $r_{1n} / r_{0n}$, with an assumed small two neutron detection probability is shown in Figure \ref{fig:cu77_p2nvsnoise_fit}.
Because it is energetically impossible for \textsuperscript{77}Cu to emit two neutrons, where the $P_{2n}$ curve crosses zero gives the estimated $r_{1n} / r_{0n}$ ratio.
This technique gives consistent results for $r_{1n} / r_{0n}$ for other nuclei that have zero $P_{2n}$ that were measured with BRIKEN.
The two neutron background coincidence rate is of order $(r_{1n}/r_{0n})^{2}$ and therefore in general can be neglected compared to the one neutron coincidence rate, though in the equations below it is tracked for the sake of completeness.

\subsection{Parent-Daughter $\beta$ Efficiencies}

The daughter nuclei may have a different $\beta$-trigger efficiency than the parent decay.
If the daughter nuclei decay has a different $\beta$-trigger efficiency than the parent nuclei decay and it is not accounted for in the Bateman equation, this will influences the fit of the parent activity.
For many decays the parent and daughter nuclei have radically different $\beta$-decay energy windows, $Q_{\beta}$ and they may have different low energy $\gamma$ rays that have large conversion electron branches. 
Both of these factors can lead to different $\beta$-detector efficiencies for parent and daughter nuclei which depend strongly on the low energy threshold of the implant-$\beta$ detector.
The Bateman equations need to be adapted in order to account for these effects and to minimize the influence of related uncertainties on $P_{xn}$.

Hereafter the $\beta$ efficiency for $P_{0n}$ decays is denoted by $\varepsilon_{\beta}$, while the $\beta$ efficiencies for $P_{1n}$ and $P_{2n}$ are given by $\varepsilon_{\beta1}$ and $\varepsilon_{\beta2}$, respectively.
The 0 in $\varepsilon_{\beta}$ is suppressed to distinguish it from the neutron multiplicity dependent $\beta$ efficiencies.

\subsection{Neutron Multiplicity Dependent $\beta$ Efficiencies}

Analogously to parent and daughter nuclei possibly having different $\beta$-detection efficiencies, the different neutron multiplicity components of a single $\beta$ decay can have different $\beta$ detection efficiencies.
The component of the $\beta$-decay with no neutrons emitted has in general a larger decay energy, $Q_{\beta}$, available for the $\beta$ and $\bar{\nu}_{e}$ to share, than for the one neutron component of the $\beta$-decay. 
This impacts the $\beta$-detection efficiency of the $\beta$ detector.
Similarly, the component of the $\beta$-decay with one neutron emitted generally has a larger decay energy, $Q_{\beta n}=Q_{\beta}-S_{n}$, available than two neutron component of the $\beta$-decay decay, $Q_{\beta 2n}=Q_{\beta}-S_{2n}$, which again can impact the $\beta$-detection efficiency. 

Another effect that impacts the $\beta$ efficiency is the final depth that the implanted nuclei stops within the implant-$\beta$ detector.
For nuclei stopped very near the silicon surface approximately 50\% of the emitted electrons leave no energy deposit in the ion-implant pixel of the $\beta$ detector.
The implantation depth also influences the number of detected minimally ionizing $\beta$ particles, which to a good approximation are $\beta$ particles with energy above 1 MeV.
Minimally ionizing $\beta$ particles deposit about 400 keV per mm of silicon.
With a $\beta$-detection threshold of 200 keV, it is possible for a high energy $\beta$ to leave less than the threshold energy in the implant-$\beta$ detector if it travels through less than $0.5$ mm of silicon.
To a first approximation to calculate the effect of the implantation depth on the $\beta$ efficiency one can assume $\sim55\%$ of minimally ionizing $\beta$s are detected.
The number of minimally ionizing $\beta$ particles can be estimated by assuming a Gamow-Teller $\beta$ emission spectrum with end-point $Q_{\beta}$, $Q_{\beta n}$, or $Q_{\beta 2n}$, as appropriate.

In this paper the $\beta$ efficiency for $\beta$ decays that emit no neutrons is written as $\varepsilon_{\beta}$, while the $\beta$ efficiencies for $\beta$ decays that emit one or two neutrons are given by $\varepsilon_{\beta1}$ and $\varepsilon_{\beta2}$, respectively.
For \textsuperscript{77}Cu ($Q_{\beta n} = 5.61$ MeV and $Q_{\beta} = 10.17$ MeV \cite{Wang2016}), an implant-$\beta$ detector threshold of 200 keV and assuming a Gamow-Teller $\beta$ distribution leads to a $\sim 1\%$ relative difference in the number of $\beta$s detected.
And, still assuming a Gamow-Teller $\beta$ distribution, up to a $\sim10\%$ relative difference in the number of high energy $\beta$ particles detected if the ion-implant position in the silicon detector is taken into account.
To account for possible additional effects, a $15\%$ uncertainty in the ratio of the one neutron emission $\beta$ efficiency to the zero neutron $\beta$ efficiency is assumed for \textsuperscript{77}Cu to be $\varepsilon_{\beta1}/\varepsilon_{\beta}=1.00(15)$.

\subsection{Energy Dependence of Neutron Efficiency}

As emphasized in \cite{belen2016}, the overall neutron efficiency depends on the energy of the emitted neutron.
The energy of neutrons emitted in $P_{(x+1)n}$ events in general will have lower energy compared with $P_{xn}$ events, though how much lower is challenging to estimate.
By using $Q_{\beta}$ and the neutron separation energy, $S_{n}$, values, estimates of the absolute upper emitted neutron energies can be made.

\section{Bateman Equations}

\subsection{Impact on Bateman Equations}

The impact of differing parent-daughter $\beta$ efficiencies is not included in the original Bateman equation solution \cite{bateman1910}.
In order to properly fit the full Bateman equation, the $P_{xn}$ need to be known, and for unmeasured $\beta$-delayed neutron emitting nuclei this is not the case.
In addition, the parent and daughter $\beta$ efficiencies need to be known.
The modification to the Bateman equation for differing parent-daughter $\beta$ efficiencies is similar to the correction due to the $P_{xn}$ daughter-neutron daughter factor, and disentangling these two values is not well defined from the fit of the adapted Bateman equation to the data.

The Bateman equation solutions for zero, one, and two neutron ion-implant $\beta$ activities depend on the $P_{xn}$ values, the parent and daughter $\beta$ efficiencies, and on the neutron efficiency in a more intricate way than the full ion-implant $\beta$-decay time activity does. 
Effectively, these parameters are not uniquely identifiable from the fit. 
Fortunately, precise knowledge of these parameters is not required to estimate the $P_{xn}$.
Even with ambiguity in the parameter values, the estimated initial activities from the neutron-multiplicity ion-implant-$\beta$ activities can be used to calculate the $P_{xn}$.

In order to minimize the influence of the relative daughter $\beta$ efficiencies and the unknown $P_{xn}$ values on the Bateman fits, the estimated initial activity of the zero, one, and two coincident neutron implant-$\beta$ activity curves ($A_{0}, A_{1}, A_{2}$) can be extracted instead of the full number of counts obtained from a original Bateman equation fit.
The initial activity precision is affected by the statistics, but is mainly influenced by the parent half-life uncertainty.
It is worth noting that the full statistics are used to estimate the initial activity.
The influence of unknown daughter $\beta$ efficiencies and of the initially unknown $P_{xn}$ dominate the errors.
The impact of these uncertainties are minimized by looking at the estimated initial activity, see figures \ref{fig:cu77_0n_fit}, \ref{fig:cu77_1n_fit}, \ref{fig:cu77_2n_fit}.
Finally, it is worth noting that the initial activity at the implant time can be read directly from the decay curve in order to make online estimates of the $P_{xn}$.

\subsection{Bateman Fitting Ranges}

The time range used for fitting the adapted Bateman equations is an important factor.
For the BRIKEN implant-$\beta$ detectors there was electronic noise in AIDA for the first 30 ms immediately after the ion-implant time, so this early time data is not included in the fit.
This noise has been corrected after the first experimental runs and the initial cutoff time has been reduced to around $10$ ms.
This electronic noise is much longer than, and therefore dominates, the ion-implant exclusion time, $T_{th}$, mentioned previously.
In the \textsuperscript{77}Cu data we do not use the first $40$ ms of data, which does not impact the calculations due to the much longer \textsuperscript{77}Cu half life of $468(2)$ ms \cite{Wang2016}.
For much shorter half lives this becomes a limiting factor.

Choosing the higher time cutoff depends on several factors.
First is the limitation of the background being modeled as linear, as discussed previously.
The second limitation is the accuracy of the modified Bateman equation and what is actually being fit as the maximum time is increased.
There is effectively no more direct information about the parent decay after six parent half lives, so fitting beyond that only gains information on the daughter and grand daughter decays.
But the daughter decays are not the primary information we are after, we are after the parent decay information.
For all of the adapted Bateman equation fits, the endpoint of each fit is varied from 6 to 10 times the parent half life.

\subsection{Connecting Multi-Neutron Activities to Neutron Branching Fractions}

The fundamental equation that contains only implant-$\beta$ time dependent terms can be written as
\begin{equation}
\left(
\begin{array}{c}
  A_{0}(t) \\
  A_{1}(t) \\
  A_{2}(t) \\
\end{array}
\right)
=
A(t) \epsilon_{I} \varepsilon_{\beta} r_{0n}
\mathbf{E}
\left(
\begin{array}{c}
  P_{0n} \\
  P_{1n} \\
  P_{2n} \\
\end{array}
\right),
\label{eq:PtoA}
\end{equation} 
where $A_{x}(t)$ is the implant-$\beta$ activity with detecting $x$ neutrons at time t (or summed over a range of times), $A(t)$ is the overall activity over the same time, $\epsilon_{I}$ is the implant efficiency, $\varepsilon_{\beta}$ is the $\beta$ efficiency for zero neutron decays, $r_{0n}$ is the probability to detect no background neutrons in a given time window, $P_{xn}$ is the branching probability for emitting $x$ neutrons, and $\mathbf{E}$ is a matrix given by 
\begin{equation}
\mathbf{E} = 
\left(
 \begin{array}{ccc}
  1 & a_{1} \epsilon_{10n} & a_{2} \epsilon_{20n} \\
  r_{1n}/r_{0n} & a_{1} \left( \epsilon_{11n} + \epsilon_{10n} r_{1n}/r_{0n} \right) & 
  a_{2} \left( \epsilon_{21n} + \epsilon_{20n} r_{1n}/r_{0n} \right) \\
  r_{2n}/r_{0n} & a_{1} \left( \epsilon_{11n} r_{1n}/r_{0n} + \epsilon_{10n} r_{2n}/r_{0n} \right) & 
  a_{2} \left( \epsilon_{22n} + \epsilon_{21n} r_{1n}/r_{0n} + \epsilon_{20n} r_{2n}/r_{0n} \right) 
 \end{array}
\right).
\label{eq:E}
\end{equation}
In the matrix $\mathbf{E}$, $a_{x}$ is the ratio of the $x$-neutron $\beta$ efficiency ($\varepsilon_{\beta x}$) to  $0$-neutron $\beta$ efficiency ($\varepsilon_{\beta}$), $\epsilon_{xyn}$ is the probability to detect $y$ neutrons given that $x$ neutrons were emitted $\left(x \geq y\right)$, and $r_{xn}$ is the probability that $x$ background neutrons are detected within a given time window. 
The matrix $\mathbf{E}$ is easily extended to include  $A_{3n}$, $A_{4n}$, $P_{3n}$, and $P_{4n}$ terms if needed.
A derivation of this fundamental equation is presented in \ref{App:derivation}.

After solving equation \ref{eq:PtoA} for the $P_{xn}$ and taking the ratio of $P_{xn}$ while requiring the sum to be 1.0, the dependence of the results on the variables $A(t)$, $\epsilon_{I}$, $\varepsilon_{\beta}$, and $r_{0n}$ is removed. 

\subsection{Initial Activity Contamination by Daughter Activities}

The early ion-implant-$\beta$ activities for the $A_{xn}(t)$ have small quantifiable contributions from the daughter decays.
By looking at early times, times much smaller than the daughter half life just after the ion-implant time, the amount of daughter activity at time $t$ is given approximately by
\begin{equation}
A_{D}(t)
\sim
\left( \lambda_{D} t \right) A_{P0},
\label{eq:a0cont}
\end{equation} 
where $A_{D}(t)$ is the daughter activity at time $t$, $\lambda_{D}$ is the daughter decay rate, and $A_{P0}$ is the initial activity of the parent.
This approximation is valid as long as $\lambda_{D} t \ll 1$ and that there are enough  $A_{P0}$ counts at early times.
In the \textsuperscript{77}Cu example, the number of daughter decays at time $t=10$ ms amounts to $\sim0.2\%$ of the initial activity of \textsuperscript{77}Cu.

\subsection{Influence of Daughter Parameters on Initial Activities}

All of the parameters related to the daughter decays, $P_{xn}$ values, daughter $\beta$ efficiencies, and daughter half lives, minimally influence the initial activity deduced from the fit.
This is because all of the parameters in the modified Bateman equation at early times are proportional to terms shown in equation \ref{eq:a0cont}.
And therefore as time goes to zero, the direct influence of the parameter uncertainties on the initial activity fit also goes to zero.
The daughter parameters still influence the estimation of the parent half life, but as we demonstrate below this error has reduced influence on the $P_{xn}$.

This line of argument is only true for experiments with no directly implanted daughter nuclei in the same pixel within the analysis time window.
For experiments with a nonzero initial daughter activity equation \ref{eq:a0cont} does not apply and hence the propagation of errors in the daughter nuclei parameters do not necessarily reduce to zero as in equation \ref{eq:a0cont}.

\subsection{Influence of Half Life on the Initial Activities}

The parent half life uncertainty influences the $P_{xn}$ uncertainty, but the impact on the calculated $P_{xn}$ is mitigated by the linear nature of the solution of equations \ref{eq:PtoA} and \ref{eq:E}.
Since the parent half life is the same for all three decay components, the impact on the $P_{xn}$ errors of the half life uncertainty is minimized.

\begin{figure}[p!]
\begin{center}
\includegraphics[width=1.0\linewidth]{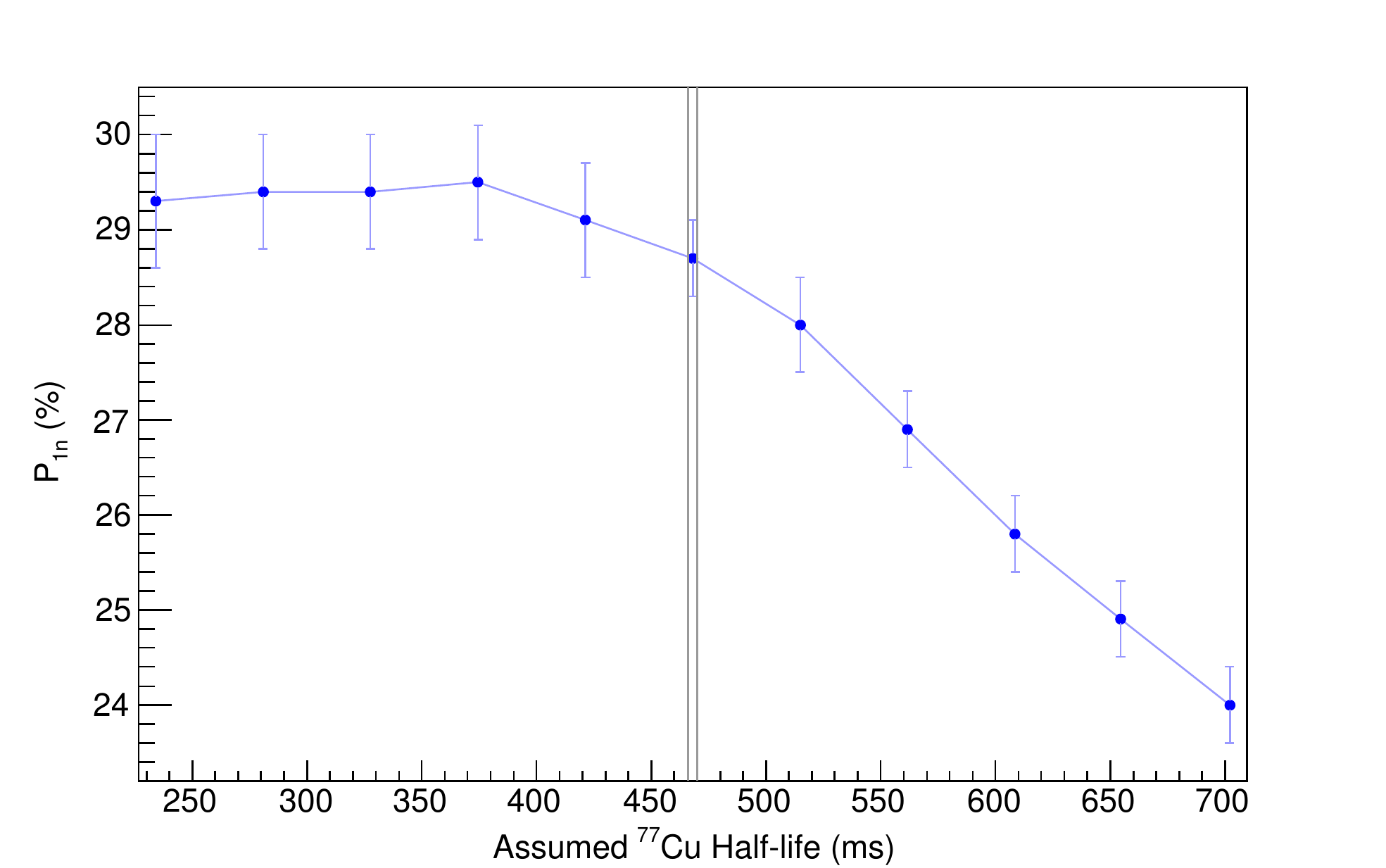}
\caption{
The variation of the calculated $P_{1n}$ versus input \textsuperscript{77}Cu half life.
This demonstrates the technique's level of stability to uncertainties in the half life.
The experimental \textsuperscript{77}Cu half life is bounded by the two gray lines \cite{Wang2016}.
}
\label{fig:cu77_p1nvshalflife}
\end{center}
\end{figure}

In figure \ref{fig:cu77_p1nvshalflife}, the assumed \textsuperscript{77}Cu half life is varied by $\pm50\%$ and the impact on the calculated \textsuperscript{77}Cu $P_{1n}$ is $(+2,-16)\%$.
If the \textsuperscript{77}Cu half life is assumed unknown by $\pm10\%$, the impact on the calculated \textsuperscript{77}Cu $P_{1n}$ is $\pm2\%$.
In the case of the literature value of \textsuperscript{77}Cu, 468(2) ms \cite{Wang2016,hosmer2009,Ilyushkin2009}, the resulting uncertainty of $P_{1n}$ is $\pm0.2\%$.
This is a negligible number when compared with the other sources of uncertainty.

One way to evaluate the half life error is to use the one neutron implant-$\beta$ activity to estimate the half life, because the uncertainty in the zero neutron implant-$\beta$ activity is usually larger.
The one neutron implant-$\beta$ activity half life is then used in the zero neutron implant-$\beta$ activity to calculate the $P_{xn}$.
We demonstrate this for the \textsuperscript{77}Cu below.
For more neutron rich nuclei, the challenge of extracting a half life due to daughter contamination will be present in the one and even the two neutron implant-$\beta$ activities and therefore it may be more challenging to obtain a precise half life.
But due to the linear nature of the ORNL BRIKEN analysis technique, the impact of the half-life error on the $P_{xn}$ is reduced.

\section{Statistical and Systematic Uncertainties Summary}

Knowledge of the parent half life has an impact on the estimated errors of $P_{xn}$.
In many cases, knowledge of the half life is available from previous experiments, but for many of the exotic neutron-rich nuclei measured with BRIKEN, the half lives are currently unknown or have extremely large uncertainties.

In $\beta$-neutron decays, up until recently it has been possible to use the one neutron decay activity to get a good half-life measurement, because it is a clean spectrum with little to no contamination from the daughter decays.
For exotic neutron-rich nuclei this may no longer be the case because the daughter nuclei decays may also have a significant $\beta$-delayed neutron decay channel, and extracting the half-life from one, and even two, neutron implant-$\beta$ activity curves may not be a precise measure of the $\beta$-decay half life.
Another effective way to measure a more precise half life is to measure an associated $\gamma$ ray and its half life gating on the $\gamma$ energy in the HPGe detectors.
But this is not always possible, such as in cases where there are no detected $\gamma$ rays associated with the particular decay, whether from low statistics or from no $\gamma$ rays being emitted.
In each case the single best possible estimate of the half life should be used to fit all of the x-neutron activity decay curves, though what is considered best will depend on the specifics of each nuclei and its daughters.

\section{Example - \textsuperscript{77}Cu}

For \textsuperscript{77}Cu the half life is well known, 468(2) ms \cite{Wang2016,hosmer2009,Ilyushkin2009}, but as an exercise, the evaluation is also presented as if the half life is unknown and the half lives for the zero, one and two neutron decay activities are treated as independent.
This means the half lives are (slightly) different for each $x$ ($x=0,1,2$) neutron implant-$\beta$ activity, which in turn leads to large uncertainties in the calculated $P_{xn}$ values.
In the analysis of nuclei measured with BRIKEN, the same half life is used for zero, one, and two neutron decay activity curves.

By varying the initial activities, $A_{x}$, with the uncertainties from the adapted Bateman equation fit and propagating the results through equation \ref{eq:PtoA} the statistical errors in the $P_{xn}$ can be calculated. 
To calculate the systematic errors, one can vary the parameters ($\epsilon_{11n}$, $a_{x}$, $r_{xn}/r_{0n}$, etc..) in equation \ref{eq:PtoA} by their respective uncertainties independently while evaluating the $P_{xn}$ repeatedly.

The decay of \textsuperscript{77}Cu is well characterized, $[ T_{1/2} = 468(2)$ ms, $Q_{\beta} = 10.17(15)$ MeV, $Q_{\beta n} = 5.61(15)$ MeV, $Q_{\beta 2n} = -2.21(15)$ MeV$]$ \cite{Wang2016}.
The negative $Q_{\beta 2n}$ for \textsuperscript{77}Cu means that two neutron decay is not possible.
In figures \ref{fig:cu77_0n_fit}, \ref{fig:cu77_1n_fit}, and \ref{fig:cu77_2n_fit} the implant-$\beta$ activities with zero, one, and two neutron multiplicity as a function of time, $A_{x}(t)$, for \textsuperscript{77}Cu are shown. 
Approximate initial activities, $A_{x}$, can be read off the histograms, though associating a precise uncertainty for the read off initial activity poses challenges.
The initial activities and uncertainties from the fits with the adapted Bateman equation without using information on the \textsuperscript{77}Cu half life and not requiring the zero, one, and two neutron implant-decay curve half lives to be the same are $A_{0}=914(106)$, $A_{1}=209(15)$, and $A_{2}=2.5(7)$. 

The initial activities and uncertainties from the fits with the adapted Bateman equation assuming the known half life, $T_{1/2}=468$ ms, are $A_{0}=908(11)$, $A_{1}=212(3)$, and $A_{2}=2.6(4)$. 
Notice the uncertainties are much smaller than in the unknown and independently varied half-life case.
The resulting \textsuperscript{77}Cu half life from the one neutron decay activity fit is $T_{1/2}=471(25)$ ms and if half life is used in the analysis of all three decay activity curves it gives identical results as using the known half life of $468(2)$ ms.

Since there are two neutron counts with a decay detected, one might naively think there is possibly a small two neutron decay branch.
But if one compares the initial two neutron activity to the initial one neutron activity, the ratio is a little over $0.01$, which is just the relative probability to detect a single random background neutron in the \textsuperscript{3}He detectors in our thermalization time window, $r_{1n}/r_{0n} = 0.012$. 
Using the same argument, about $10$ of the one neutron activity counts, $A_{1}=212(3)$, are actually zero neutron events in coincidence with a background neutron.
In this case it is a small correction, $\sim 5\%$ relative error, but in other cases with different relative $P_{xn}$ values this can be a much larger correction.
For example, a large $P_{0n}$ and a small $P_{1n}$, on the order of a percent or two, will have a large component of random coincidences in the one neutron decay curve.
This observation holds similarly for a large $P_{1n}$ and a small $P_{2n}$.

\begin{figure}[p!]
\begin{center}
\includegraphics[width=1.0\linewidth]{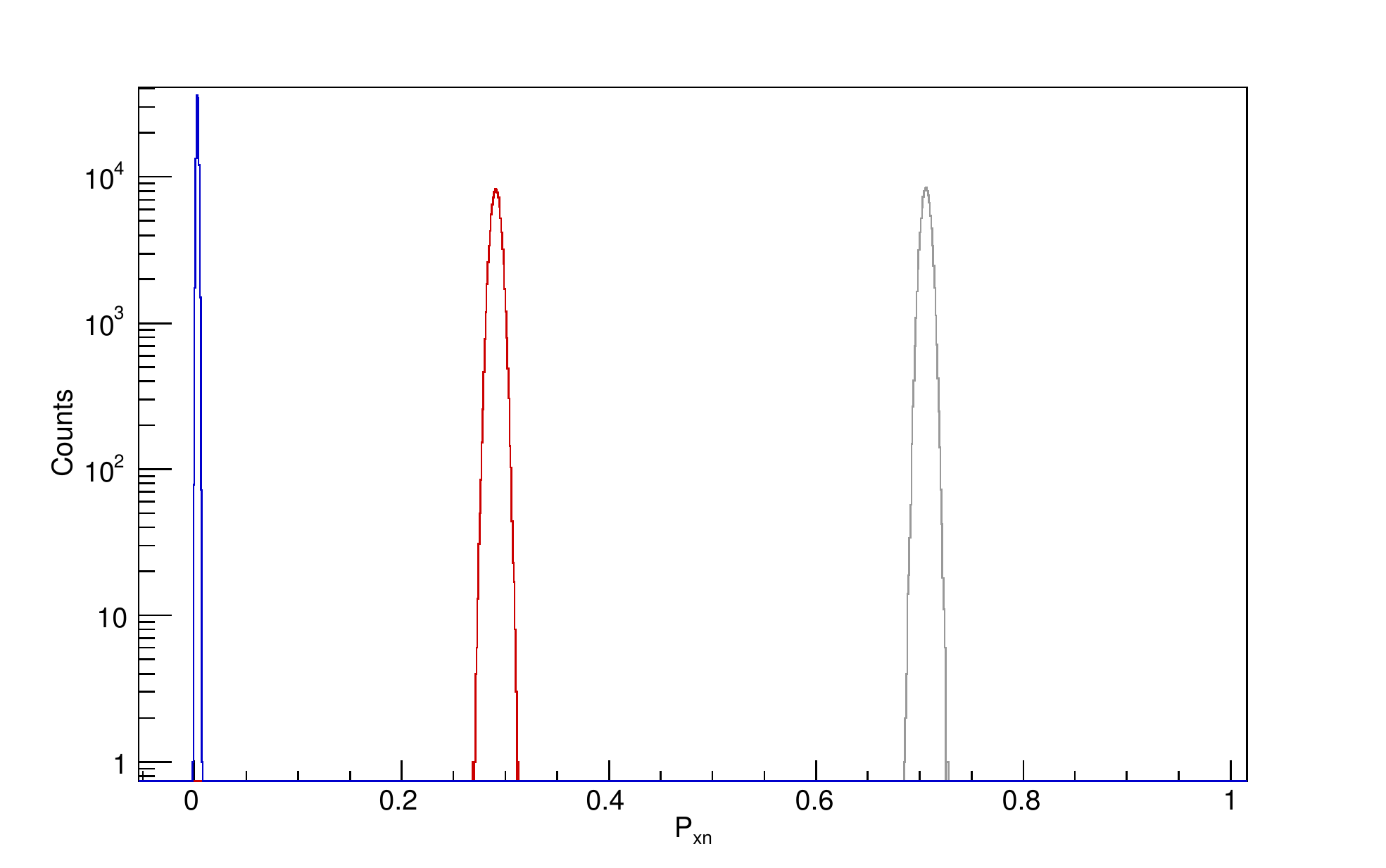}
\caption
{
Statistical variation of \textsuperscript{77}Cu initial activities and the impact on the $P_{xn}$ assuming the known \textsuperscript{77}Cu half life, $T_{1/2}=468(2)$ms.
$P_{0n}$ is shown in gray, $P_{1n}$ is shown in red, and $P_{2n}$ is shown in blue.
}
\label{fig:77cu_stat_fit_fixed_halflife}
\end{center}
\end{figure}

\begin{figure}[p!]
\begin{center}
\includegraphics[width=1.0\linewidth]{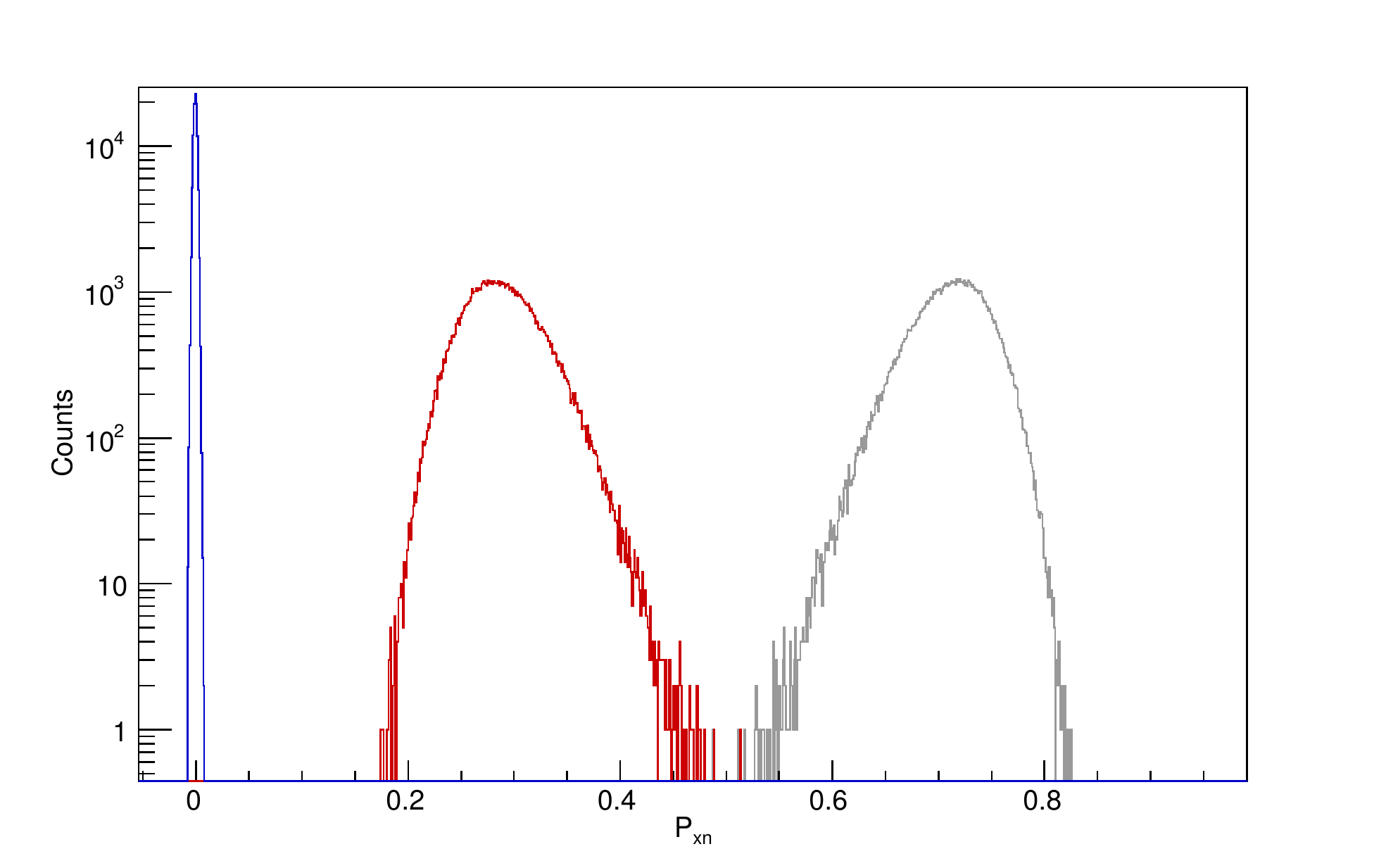}
\caption
{
Statistical variation of \textsuperscript{77}Cu initial activities and the impact on the $P_{xn}$ with non-fixed \textsuperscript{77}Cu half life.
Colors are as in Figure \ref{fig:77cu_stat_fit_fixed_halflife}.
}
\label{fig:77cu_stat_fit}
\end{center}
\end{figure}

Using these initial activities and assuming a single neutron efficiency of $62\%$ \cite{tar2017}, a relative daughter $\beta$ efficiency, $a_1=1.0$, and estimating the noise by requiring the $P_{2n}$ is zero which gives $r_{1n}/r_{0n}=0.012$, as shown in figure \ref{fig:cu77_p2nvsnoise_fit}.
For the case where the \textsuperscript{77}Cu half life is fixed to the known value and varying the $A_{x}$ by their uncertainties 100,000 times while inputing these values into equation \ref{eq:PtoA}, a fit of the resulting distribution is shown in figure \ref{fig:77cu_stat_fit_fixed_halflife} with a Gaussian function and reporting the $\bar{P}$ and $\sigma_{P}$, one obtains $P_{0n}=71.2(5)\%$, $P_{1n}=28.8(5)\%$, and $P_{2n}=0.000(1)\%$. 
For the case with an unconstrained \textsuperscript{77}Cu half life and the same neutron efficiency one obtains $P_{0n}=71.1(33)\%$, $P_{1n}=28.9(33)\%$, and $P_{2n}=0.000(2)\%$, the results are shown in figure \ref{fig:77cu_stat_fit}. 

\begin{figure}[p!]
\begin{center}
\includegraphics[width=1.0\linewidth]{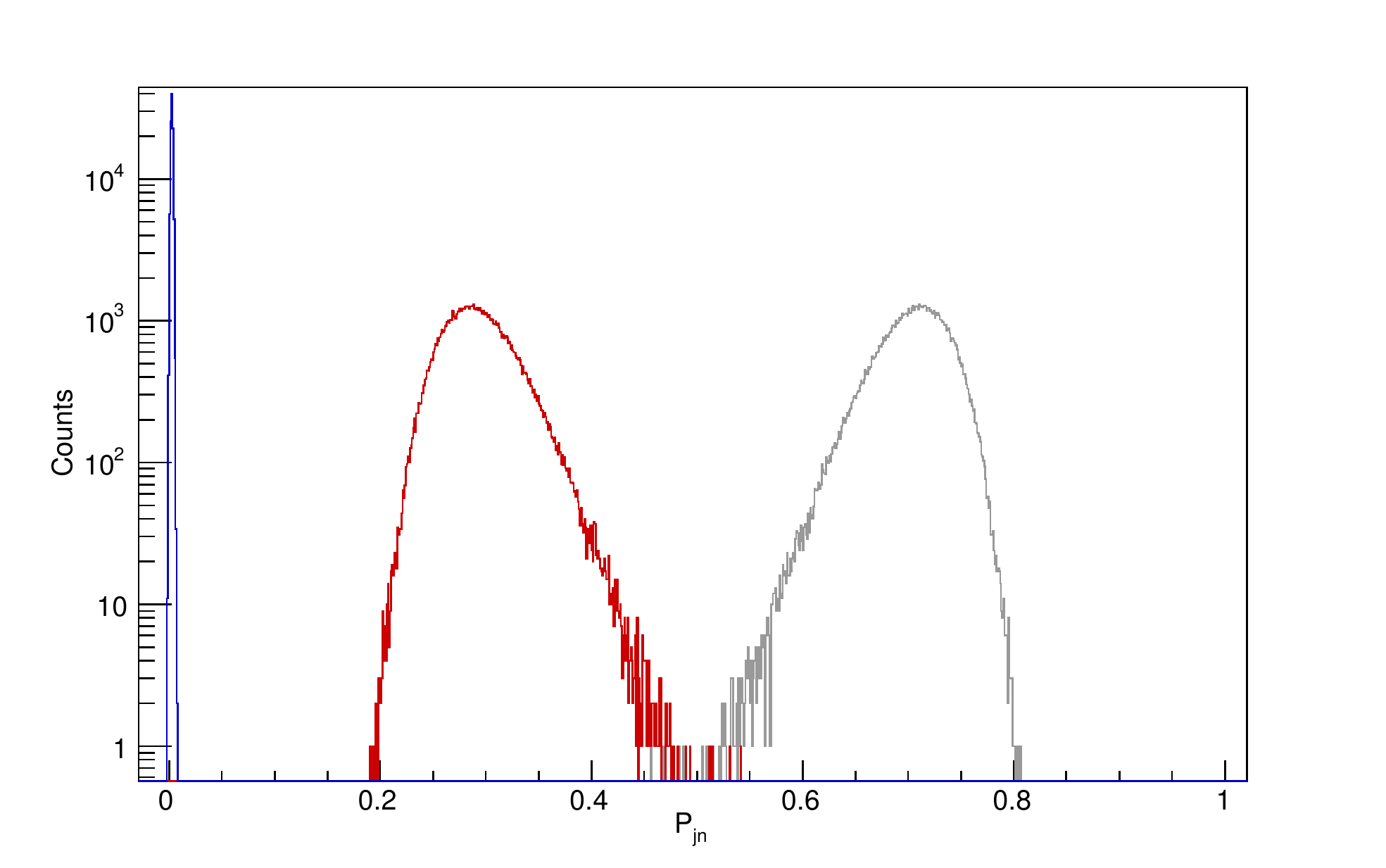}
\caption
{
Statistical and systematic errors after variation of \textsuperscript{77}Cu initial activities and the other parameters described in the text and their impact on the $P_{xn}$ assuming the known \textsuperscript{77}Cu half life, $T_{1/2}=468(2)$ms.
Colors are as in Figure \ref{fig:77cu_stat_fit_fixed_halflife}.
}
\label{fig:77cu_total_fit_fixed_halflife}
\end{center}
\end{figure}

\begin{figure}[p!]
\begin{center}
\includegraphics[width=1.0\linewidth]{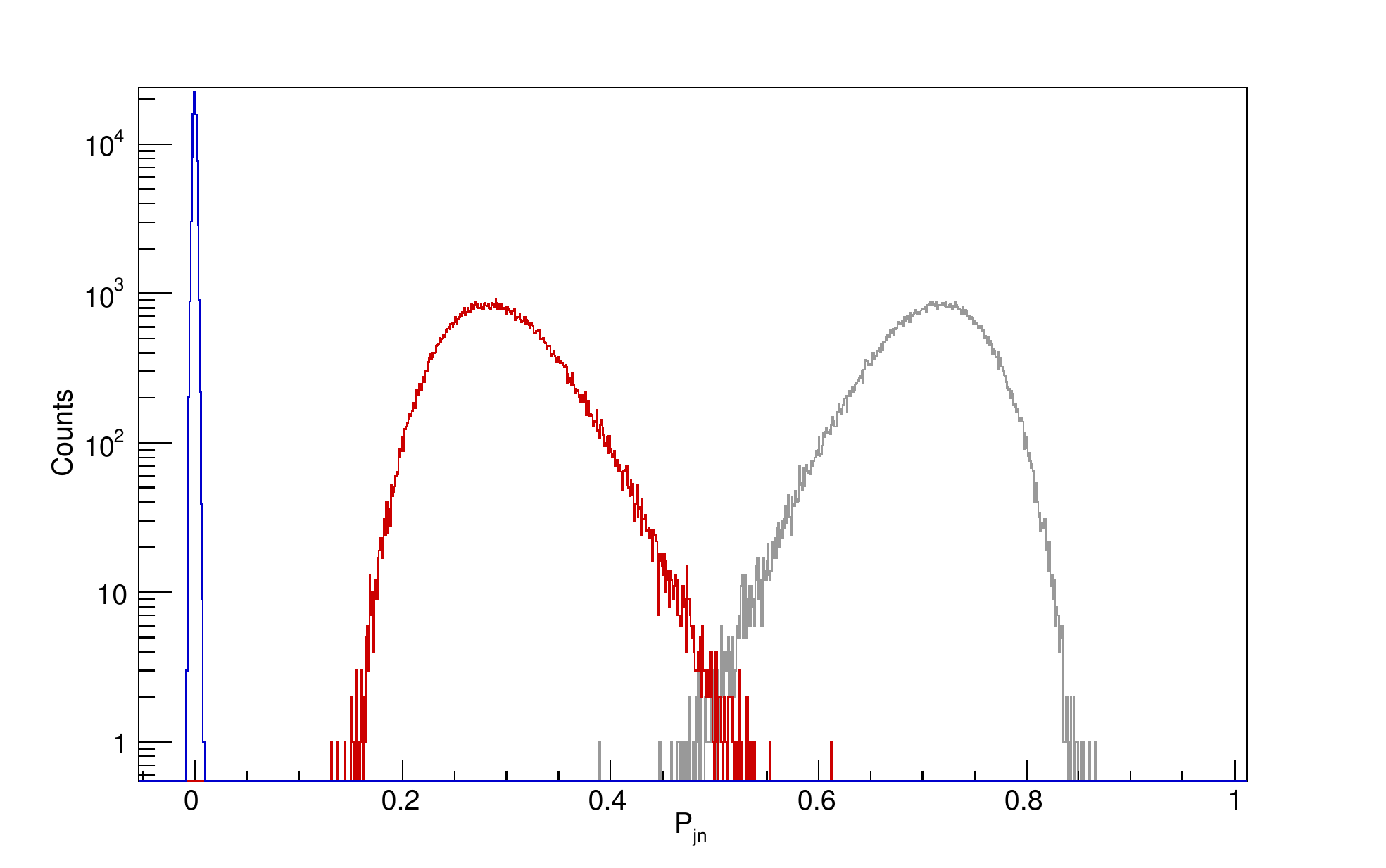}
\caption{
Systematic and statistical variation of \textsuperscript{77}Cu initial activities and the other parameters described in the text and their impact on the $P_{xn}$ with a non-fixed \textsuperscript{77}Cu half life.
Colors are as in Figure \ref{fig:77cu_stat_fit_fixed_halflife}.
}
\label{fig:77cu_total_fit}
\end{center}
\end{figure}

If in addition to the statistical uncertainties, the single neutron efficiency is varied as $62(2)\%$ \cite{tar2017}, and the relative neutron-multiplicity as $\beta$ efficiency as $a_1=1.00(15)$ (motivated previously), the calculated $P_{xn}$ distributions are shown in figures \ref{fig:77cu_total_fit_fixed_halflife} and \ref{fig:77cu_total_fit}.
Fitting each distribution with a Gaussian function, one obtains $P_{0n}=70.8(30)\%$, $P_{1n}=29.2(30)\%$, and $P_{2n}=0.000(1)\%$ using the known half life and leaving the half life unconstrained one obtains $P_{0n}=70.7(44)\%$, $P_{1n}=29.3(44)\%$, and $P_{2n}=0.000(2)\%$.  

Since the \textsuperscript{77}Cu half life is well known, our reported one neutron branching fraction, $P_{1n}=29.2(30)\%$, is in 1 $\sigma$ agreement with the literature values of $P_{1n}=31.0(38)\%$ \cite{hosmer2009} and $P_{1n}=30.3(22)\%$ \cite{Ilyushkin2009}.
The two literature values were obtained using two different techniques, providing confidence in the value.

\section{Summary}

We have presented the fundamentals of the BRIKEN analysis and shown two evaluations of \textsuperscript{77}Cu $\beta$-neutron precursor decay properties and their associated statistical and systematic uncertainties as examples.
We present a general result that simplifies calculation and propagation of uncertainties, this is shown in equations \ref{eq:PtoA} and \ref{eq:E}.
We also present a discussion of extracting zero, one, and two neutron activities appropriate for the BRIKEN implant-$\beta$ trigger setup. 
This discussion is applicable to other experiments if daughter implants are spatially and temporally distinguishable from the nuclei of interest implants.
If this is not an appropriate description of a particular other experiment, the conversion of activities to $P_{xn}$ in equations \ref{eq:PtoA} and \ref{eq:E} is still valid.
For \textsuperscript{77}Cu the BRIKEN result for the one neutron branching fraction, $P_{1n}=29.2(30)\%$ agrees with previous measurements of $P_{1n}$ in the literature. 
This agreement increases our confidence in the evaluation procedure presented in this paper.

\section{Acknowledgements}
This research was sponsored by the Office of Nuclear Physics, U. S. Department of Energy under contracts DE-AC05-00OR22725 (ORNL).
This work was also supported in part by the National Science Foundation grant PHY 1714153 (CMU).
This work was supported by the Spanish Ministerio de Econom\'ia y Competitividad under Grant, No. FPA2014-52823-C2-1-P,  and the program Severo Ochoa (SEV-2014-0398)
This work was also supported and inspired by the IAEA Coordinated Research Project for a ``Reference Database for $\beta$-Delayed Neutron Emission''.

\bibliography{BRIKEN_Bibliography}





\appendix
\section{Derivation of Equations \ref{eq:PtoA} and \ref{eq:E}}
\label{App:derivation}
In this appendix we describe the derivation of equation \ref{eq:PtoA} and \ref{eq:E}.
For the derivation we only consider up to a two neutron emitting nucleus.
The extension of the analysis to three and four neutron decays is straight forward.
The basis of the derivation is to consider all of the possible ways to detect y neutrons ($0 \leq y \leq x$) given that x neutrons ($0 \leq x \leq 2$) are emitted.
For clarity, in the first part of the derivation we ignore the dependence of the relative $\beta$ efficiency on the number of neutrons emitted, that modification is shown following the basic derivation.

The possible ways to detect no neutrons for various decay events are listed here.
There are only three possible ways.
The first possibility is a decay with zero neutrons emitted and no background neutrons detected.
The second possibility is a decay with one neutron emitted but that neutron is not detected and no background neutrons are detected.
The third possibility is a decay with two neutrons emitted but neither neutron is detected and no background neutrons are detected.
Using the notation used in equations \ref{eq:PtoA} and \ref{eq:E}, the ways to detect zero neutrons can be written as
\begin{equation}
A_{0}(t) = A(t) \epsilon_{I} \varepsilon_{\beta} r_{0n} \left(P_{0n} +  \epsilon_{10n} P_{1n} + \epsilon_{20n} P_{2n} \right).
\label{eq:a0activity}
\end{equation} 

Next is the list of possible ways to detect one neutron from various decay events.
There are five possible ways.
The first possibility is a decay with zero neutrons emitted and one background neutron detected.
The second possibility is a decay with one neutron emitted and that neutron is detected and no background neutrons are detected.
The third possibility is a decay with one neutron emitted but that neutron is not detected and one background neutron is detected.
The fourth possibility is a decay with two neutrons emitted and only one of those neutrons are detected and no background neutrons are detected.
The fifth possibility is a decay with two neutrons emitted and neither of those neutrons are detected but one background neutron is detected.
Using the notation used in equations \ref{eq:PtoA} and \ref{eq:E}, the ways to detect one neutron can be written as
\begin{equation}
A_{1}(t) = A(t) \epsilon_{I} \varepsilon_{\beta} \left(P_{0n} r_{1n} +  \epsilon_{11n} r_{0n} P_{1n} +  \epsilon_{10n} r_{1n} P_{1n} + \epsilon_{21n} r_{0n} P_{2n} + \epsilon_{20n} r_{1n} P_{2n} \right),
\label{eq:a1activity}
\end{equation} 

The last enumeration of possibilities considered is the list of possible ways to detect two neutrons from various decay events.
There are six possible ways.
The first possibility is a decay with zero neutrons emitted and two background neutron detected.
The second possibility is a decay with one neutron emitted and that neutron is detected in coincidence with one background neutron detected.
The third possibility is a decay with one neutron emitted but that neutron is not detected but two background neutrons are detected.
The fourth possibility is a decay with two neutrons emitted and both emitted neutrons are detected along with no background neutrons detected.
The fifth possibility is a decay with two neutrons emitted and only one of the emitted neutrons is detected along with one background neutron detected.
Lastly, the sixth possibility is a decay with two neutrons emitted and neither of the emitted neutrons is detected but two background neutrons are detected.
Using the notation for equations \ref{eq:PtoA} and \ref{eq:E}, the ways to detect two neutrons can be written as
\begin{equation}
A_{2}(t) = A(t) \epsilon_{I} \varepsilon_{\beta} \left(P_{0n} r_{2n} +  \epsilon_{11n} r_{1n} P_{1n} +  \epsilon_{10n} r_{2n} P_{1n}  + \epsilon_{22n} r_{0n} P_{2n} + \epsilon_{21n} r_{1n} P_{2n} + \epsilon_{20n} r_{2n} P_{2n} \right).
\label{eq:a2activity}
\end{equation} 
Equations \ref{eq:a0activity}, \ref{eq:a1activity}, and \ref{eq:a2activity} are not quite equations \ref{eq:PtoA} and \ref{eq:E}, one additional set of parameters remains to be inserted.

Due to the possible large difference between $Q_{\beta}$, $Q_{\beta n}$, and $Q_{\beta 2n}$ (decay energy for zero, one, and two neutron decays) the associated $\beta$ efficiencies ($\varepsilon_{\beta}$, $\varepsilon_{\beta1}$,$\varepsilon_{\beta2}$) may not be the same. 
Adding these parameters to the equations, the zero neutron equation becomes
\begin{equation}
A_{0}(t) = A(t) \epsilon_{I} r_{0n} \left(\varepsilon_{\beta} P_{0n} +  \varepsilon_{\beta1}  \epsilon_{10n} P_{1n} +  \varepsilon_{\beta2} \epsilon_{20n} P_{2n} \right),
\label{eq:a0activityFull}
\end{equation} 
with similar changes to the one and two neutron equations.

After factoring out $\varepsilon_{\beta}$, $r_{0n}$, and group the $A_{x}(t)$ and the $P_{xn}$ into vectors, the remaining components are the matrix \textbf{E}, we arrive at the equations \ref{eq:PtoA} and \ref{eq:E}, the basis of the ORNL BRIKEN analysis technique.

The extension of this analysis to three and larger neutron emission is straight forward, with the additional modification that the random probability of three and four background neutrons should be included and that the $\beta$ efficiencies and neutron efficiencies for three and four neutron decays should be included.

\end{document}